\documentclass[proof]{pasj00}
\draft

\begin{document}
\SetRunningHead{Author(s) in page-head}{Running Head}

\title{The image slicer for the Subaru Telescope High Dispersion Spectrograph}

\author{Akito \textsc{Tajitsu}} %
\affil{The Subaru Telescope, National Astronomical Observatory of Japan, 650 North A'ohoku Pl., Hilo, HI 96720, USA}
\email{tajitsu@subaru.naoj.org}

\author{Wako \textsc{Aoki}}
\affil{National Astronomical Observatory of Japan, 2-21-1 Osawa, Mitaka, Tokyo, 181-8588}\email{aoki.wako@nao.ac.jp}
\and
\author{Tomoyasu {\sc Yamamuro}}
\affil{OptCraft, 3-16-8-101 Higashi-Hashimoto, Midori-ku,
            Sagamihara, Kanagawa, 252-0144}\email{yamamuro@optcraft.com}

%


\maketitle

\begin{abstract}

We report on the design, manufacturing, and performance of the image
slicer for the High Dispersion Spectrograph (HDS) on the Subaru
Telescope.  This instrument is a Bowen-Walraven type image slicer
providing five 0.3 arcsec $\times$ 1.5 arcsec images with a resolving
power of $R=\lambda/\delta \lambda = 110,000$. The resulting resolving
power and line profiles are investigated in detail, including
estimates of the defocusing effect on the resolving power.  The
throughput in the wavelength range from 400 to 700~nm is higher than
80\%, thereby improving the efficiency of the spectrograph by a factor
of 1.8 for 0.7 arcsec seeing.

\end{abstract}

\section{Introduction}

Optical spectroscopy with a very high resolving power
($R=\lambda/\delta \lambda > 100,000$) enables measurements of 
isotopic abundance ratios including $^{7}$Li/$^{6}$Li,
$^{25}$Mg/$^{24}$Mg, $^{151}$Eu/$^{153}$Eu for stellar atmospheres
and/or interstellar matter based on detailed analyses of spectral line
profiles (e.g., Smith et al. 1993; Kawanomoto et al. 2009; Aoki et
al. 2003). These studies also require very high signal-to-noise ratios
in general, which can be achieved by large telescopes.

One difficulty in such observations and instrumentation is that the
size of the stellar image at the focal plane increases with telescope
aperture, thereby requiring a larger spectrograph with a wider slit
width to achieve the same spectral resolution. One solution to achieve
very high resolution with high efficiency is to install an image
slicer. A brief history of the development of image slicers since
\citet{bowen38}, as well as the application to the VLT/UVES, is
provided by \citet{dekker03}.

The High Dispersion Spectrograph (HDS: Noguchi et al. 2002) on the
Subaru Telescope can achieve a very high resolving power, up to
150,000, by applying a very narrow slit. When a slit of 0.3~arcsec
(150~$\mu$m) width is applied, the resolving power is 115,000. The
throughput at the slit is, however, as low as 45~\% for the typical
seeing size of the telescope at Mauna Kea in Hawaii ($\sim$
0.6~arcsec).  In order to improve the efficiency of the spectrograph
for observations with very high resolution, we installed a
Bowen-Walraven type image slicer.  
This type of image slicer traps incident light in a thin plate by
  total internal reflection, and slices the image by sending the light
  at every second reflection to a glass prism inclined with an
  appropriate angle (see figure 1). After the designing of the
  instrument in late 2008, the instrument was constructed in 2009 and
  was installed in 2010. The instrument was opened for common-use of
  HDS from August 2011.


This paper reports on the design and manufacturing of the image slicer
for the HDS (\S~\ref{sec:design}), and its performance as verified using
calibration sources and stellar light (\S~\ref{sec:performance}). In
particular, the line profiles and spectral resolution obtained with
the image slicer are investigated in detail, and the effect of
defocusing at each sliced image is discussed. The operation and data
reduction procedure are also presented (\S~\ref{sec:operation}).

\section{Design and construction}\label{sec:design}

\subsection{Optical design}\label{sec:optics}

\begin{figure}
  \begin{center}
    \FigureFile(160mm,160mm){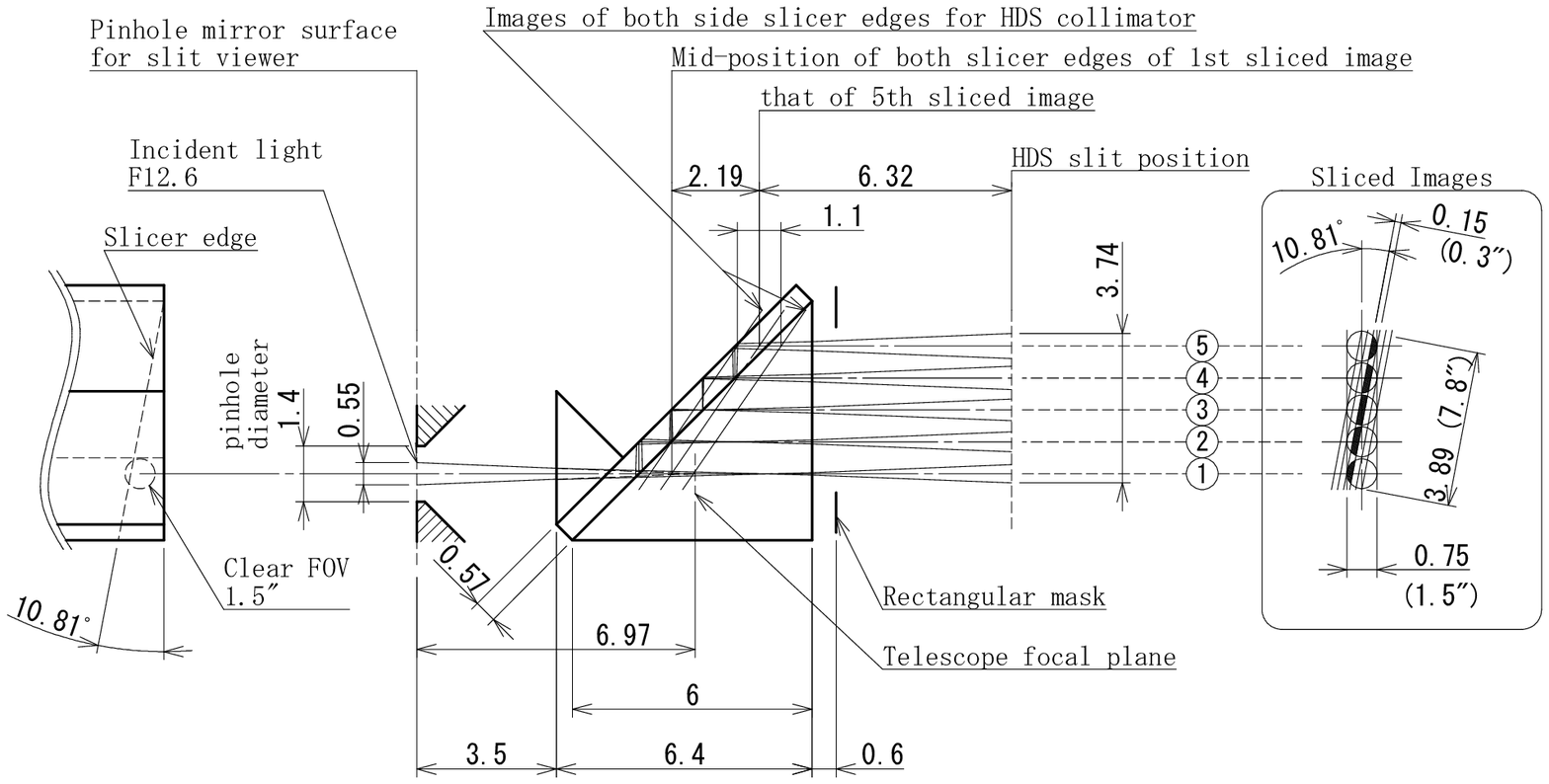} 
  \end{center}
  \caption{Optical design of the image slicer.
The light from the telescope enters from left through the pinhole
that is sufficiently large to make a clear field-of-view of 1.5 arcsec
required for observations (the clear field of view is actually 1.7
arcsec). 
The light is reflected in the thin
    (0.57~mm) plate and escapes to the inclined prism (by
    10.81~degree) glued to the plate. The right panel shows the sliced
    images obtained by this instrument.
}\label{fig:design}
\end{figure}

The Bowen-Walraven type image slicer we installed is optimized for the
following two prime requirements.  The first requirement is to
maximize the energy from point sources under the typical seeing of 0.6
arcsec on Mauna Kea. Therefore, the clear aperture should be up to 1.5
arcsec (corresponding to 0.75 mm on the focal plane of the Subaru
Telescope).  The second requirement is the spectral resolving power as
high as $R \sim 110,000$. Therefore, the width of the sliced images
should be 0.3 arcsec (0.15 mm). In order to satisfy these two
requirements, the image slicer is designed to transform a 1.5 arcsec
$\times$ 1.5 arcsec field of view into five sliced images of 0.3
arcsec width as shown in figure~\ref{fig:design}. In this paper, the
five sliced images (hereafter ``slices'') are numbered 1-5 along the
optical path as shown in the figure. The specification of the image
slicer is given in table~\ref{tab:spec}. We note that since targets of
this instrument are assumed to be bright stars, no slice is dedicated
to obtain a background sky spectrum.

Echelle spectra for the slit length of 7.8 arcsec (five 1.5 arcsec
images with small separations) are obtained without any overlap
between the adjacent orders for wavelengths longer than 4900~{\AA}
(4000~{\AA}) using the red (blue) cross disperser grating. For
instance, the wavelength range from 4900~{\AA} to 7600~{\AA} is
covered by a single exposure.


A photograph of the optical element of the image slicer is shown in
figure~\ref{fig:photo}.  The image slicer is made by optical contact
of three fused-silica pieces, which are a small triangular prism, a
parallel plate, and a large triangular prism with an inclined slicer
edge.  The reflection losses of the entrance and exit surfaces are
suppressed less than 1 \% in the wavelength range of 400 - 700 nm 
by using multilayer anti-reflection coating.

\subsection{Layout}\label{sec:holder}

The holder of the optical element is designed to mount it in front of
the HDS slit without updating of the HDS system for keeping operation
with the normal slit.  Instead of the normal slit mirror,
a pinhole mirror is placed just in front of the image slicer for
target acquisition and guiding with the HDS slit viewer. The pinhole
diameter is 2.8 arcsec (1.4 mm) and the location is 6.97~mm from the
typical slicer edge position. This makes a 1.7~arcsec (0.83 mm) clear
image without vignetting in the image slicer, and also makes a blurred
region around the clear image. The blurred region, corresponding in
general to outer region of a target star, produces a weak envelope on
one-side of the first slice, resulting in a small decrease of the
spectral resolution as described in \S~\ref{sec:th}.  Just behind the
image slicer, a 8.8 arcsec (4.4 mm) $\times$ 1.2 arcsec (0.6 mm)
rectangular mask is placed to suppress scattered light.  After the
mask, the sliced images pass through the HDS slit that is fully opened
for working not as a slit but as an open aperture.  The sliced images
are then aligned just on the center of the HDS slit position by a
mechanism connected with the holder that adjusts the x-shift, y-shit,
and position angle. In the holder, the slicer edge is located 6.32 -
8.51 mm from the HDS slit, and the HDS collimator is shifted to focus
on the image slicer instead of the normal slit (see next
section).

The above layout is different from that of the image slicer for the
VLT/UVES \citep{dekker03}. The UVES image slicer is mounted just on
the VLT focus, and sliced images are re-focused on the UVES slit by a
relay optical system. This layout makes up a sharp slit image by the
normal slit even for the sliced images, avoiding degradation of
spectral resolution by defocusing effect. On the other hand, our
layout avoids light loss due to vignetting by a slit and reflection at
relay lens surfaces. In addition, our layout simply enables us to add
the image slicer without updating of the HDS slit and the slit viewer
mechanism.

%
%
%
%
%
%


\begin{figure}
  \begin{center}
    \FigureFile(80mm,80mm){imageslicer.eps} 
  \end{center}
  \caption{Photograph of the optical element of the image slicer}\label{fig:photo}
\end{figure}

\begin{table}
  \caption{Specifications of the image slicer}\label{tab:spec}
  \begin{center}
    \begin{tabular}{ll}
      \hline
      Type             & Bowen-Walraven type \\
      Number of slices & 5  \\
      Entrance         & 1.5~arcsec $\times$ 1.5~arcsec \\
      Sliced image     & 0.3~arcsec $\times$ 7.8~arcsec \\
      Wavelength coverage & 400--700~nm \\
      Plate thickness  & 0.57~mm \\
      Material         & fused-silica \\
      \hline
    \end{tabular}
  \end{center}
\end{table}

\section{Performance of the instrument}\label{sec:performance}

\subsection{Instrument setup}\label{sec:setup}

As shown in figure~\ref{fig:design}, the individual sliced images
are located 6.32--8.51~mm from the HDS slit. These offsets are (partially)
compensated by the shift of the collimator position. In order to find
the correct position of the collimator, we first focused
the camera system using the normal slit \citep{noguchi02}. Then, we
searched for the collimator position that produces the highest spectral
resolution with the image slicer. This results in a shift of the
collimator by 7.3~mm. This well agrees with the expectation from the
design of the mount holder for the third-fourth slices.

The differences of the best focus positions for individual slices
result in degradation of spectral resolution. This problem is
discussed in the next subsection.

\begin{figure}
  \begin{center}
    \FigureFile(120mm,80mm){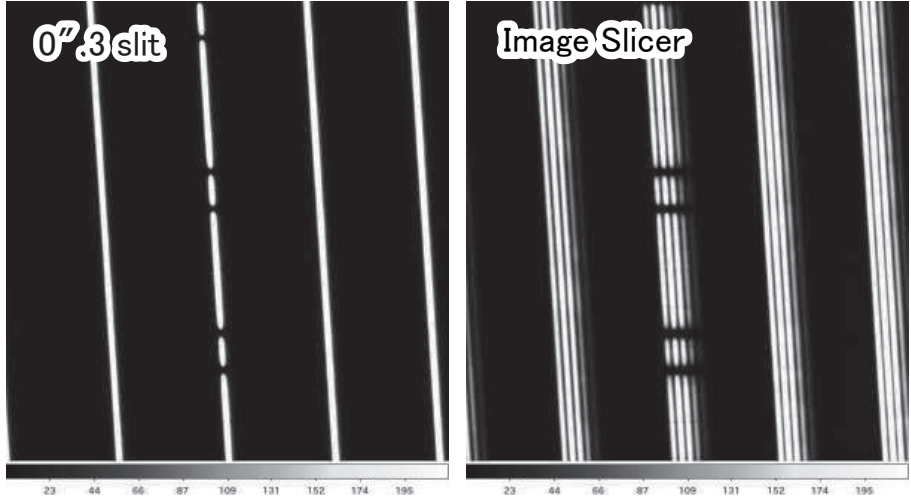} 
  \end{center}
  \caption{The CCD image of a stellar spectrum (o Per) obtained with
    the image slicer (right) and that with the normal slit with 
    0.3~arcsec width (left).}\label{fig:star2d}
\end{figure}

\begin{figure}
  \begin{center}
    \FigureFile(80mm,80mm){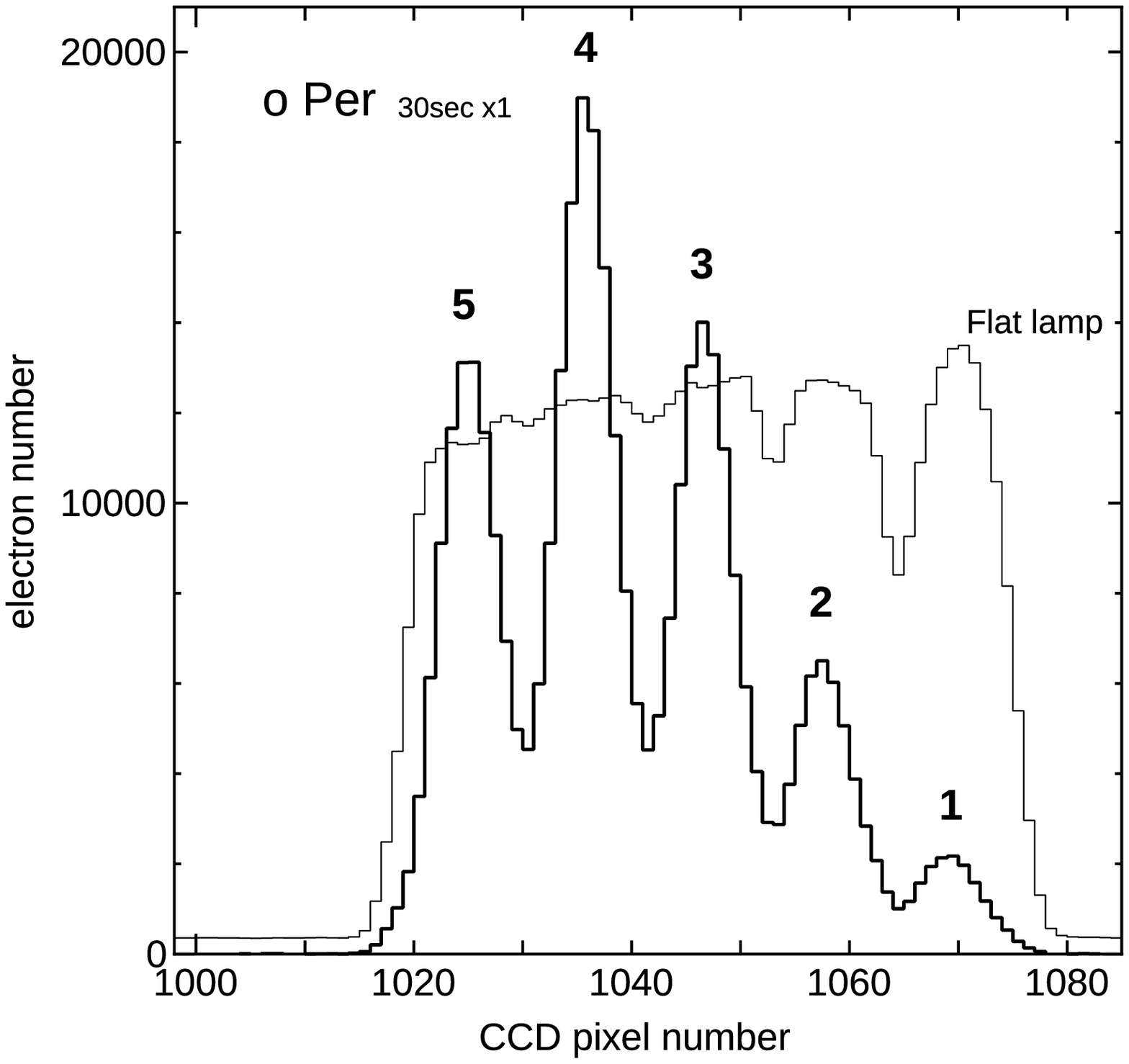} 
  \end{center}
  \caption{A cross cut image of the spectral data for an object (o
    Per) and the flat lamp obtained with the image slicer. The low
    counts between the first and second slices found in the flat lamp
    data are because the beam from the flat lamp passes through the pin
    hole, which is in front of the image slicer, and the first slice is 
    not fully illuminated (see
    Figure~\ref{fig:design}).}\label{fig:cut}
\end{figure}

The spectral format (spectral images on the detector) obtained with
the image slicer is set to be the same as that obtained with the
normal slit by applying a small offset to the cross disperser grating
angle. As a result, direct comparisons of spectral images obtained
with and without the image slicer are possible.
Figure~\ref{fig:star2d} shows stellar spectral images (o Per) obtained
with the image slicer and a normal slit with a 0.3~arcsec width. In
the central spectrum of each panel, absorption features of telluric
lines appear. Figure~\ref{fig:cut} shows a cross cut of the CCD image
for a spectrum obtained with the slicer, in which five slices are
identified corresponding to those presented in
figure~\ref{fig:design}. The cross cut for a spectral image of the
flat lamp is also depicted for comparison purposes by the thin line.



\subsection{Spectral resolution and line profile}\label{sec:th}

\begin{figure}
  \begin{center}
    \FigureFile(120mm,80mm){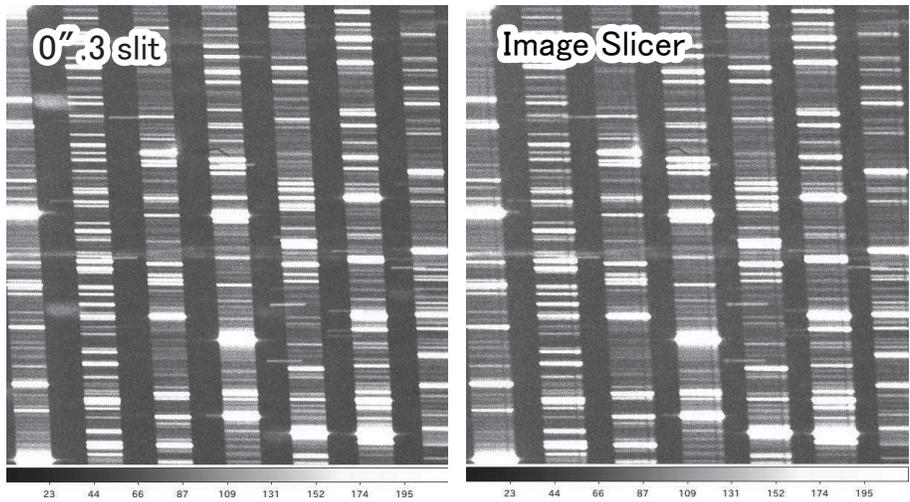} 
  \end{center}
  \caption{The same as figure~\ref{fig:star2d}, but for Th-Ar spectra.}\label{fig:thar}
\end{figure}

The spectral line profiles and the resolving power are measured for
weak emission lines of Th-Ar arc spectra. Figure~\ref{fig:thar} shows
a section of the CCD image obtained with and without the image
slicer. The image without the slicer is obtained with the slit of
0.3~arcsec (0.15~mm) width and 7.5~arcsec (3.75~mm) length.

Figure~\ref{fig:profile1} shows the spectral profiles of ten weak Th
emission lines around 670~nm with different symbols for
individual lines, which are normalized at the peak by fitting gaussian
profiles. The solid line is the gaussian profile for the average of
the full width half maximums measured from the individual lines. The
measurements are made for the five slices.  The Th-Ar spectrum
obtained with the normal slit is also separated into five spectra using the
same sub-apertures as for the extraction of spectra obtained with the
slicer.

The line profile of the central (the third) slice is shown in the
figure as an example. There is apparently no clear difference between
the resolution and the profile between the spectra with and without
the image slicer. The average of the spectral resolution, $R=
\lambda/\delta \lambda$, where $\delta \lambda$ is determined as the
FWHM of the profiles for individual lines, is typically 110,000 when
the image slicer is used. For comparisons, the resolution is 114,000
when using a 0.3~arcsec slit without the image slicer (see below for
more details).

Similar results are obtained for other slices. The exception is the
first spectrum that appears in the right edge of the spectral image in
figure~\ref{fig:thar}. The profile of the spectrum of this slice
(figure~\ref{fig:profile2}) shows some asymmetry and lower spectral
resolution ($R\sim 90,000$). This is due to the contribution of the
blurred light around the clear image formed through the pinhole on the
first slice (\S~\ref{sec:holder}).

The spectral resolution obtained at each slice is shown in
figure~\ref{fig:res}. The resolution obtained with the 0.3~arcsec slit
is approximately 114,000. The reason for the weak dependence of the
resolution on the position at the slit is unclear.


The spectral resolution obtained with the image slicer is highest at
the fourth slice, suggesting that the collimator focuses around this
position (see also \S~\ref{sec:setup}). The resolution decreases by
about one percent at the adjacent slices, and by four percent at the
second slice.  The resolution at the first slice is significantly
lower for the reasons mentioned above.


\begin{figure}
  \begin{center}
    \FigureFile(80mm,80mm){fig6a.ps} 
    \FigureFile(80mm,80mm){fig6b.ps} 
  \end{center}
  \caption{Line profiles of 10 Th lines around 670~nm obtained with
    the third (central) slice (left) and that of the center of the
    slit without the slicer (right). Individual lines are normalized
    at the peak by gaussian fitting and shown by data points with
    different symbols. The solid line indicates the gaussian profile
    for the average of the full width half maximums measured
    for individual lines. There are no significant differences in the width and
    profile between the two spectra.}\label{fig:profile1}
\end{figure}
\begin{figure}
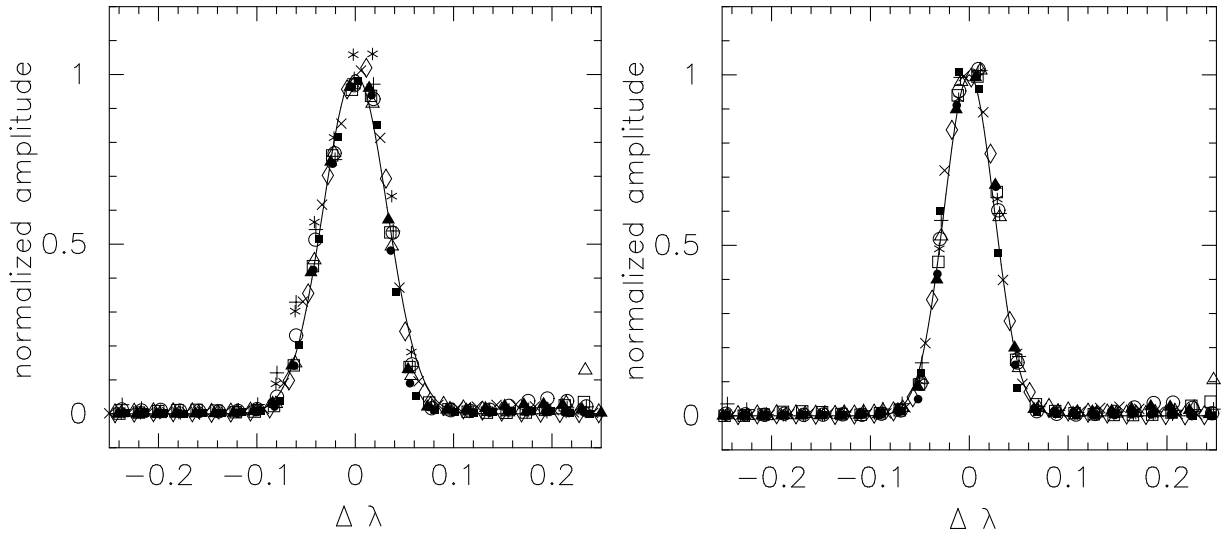

  \begin{center}
    \FigureFile(80mm,80mm){fig7a.ps} 
    \FigureFile(80mm,80mm){fig7b.ps} 
  \end{center}
  \caption{The same as figure~\ref{fig:profile1}, but for the first slice
    (left) and the corresponding data without the slicer (right). The
    spectral line of this slice is wider and shows some
    asymmetry.} \label{fig:profile2}
\end{figure}
\begin{figure}
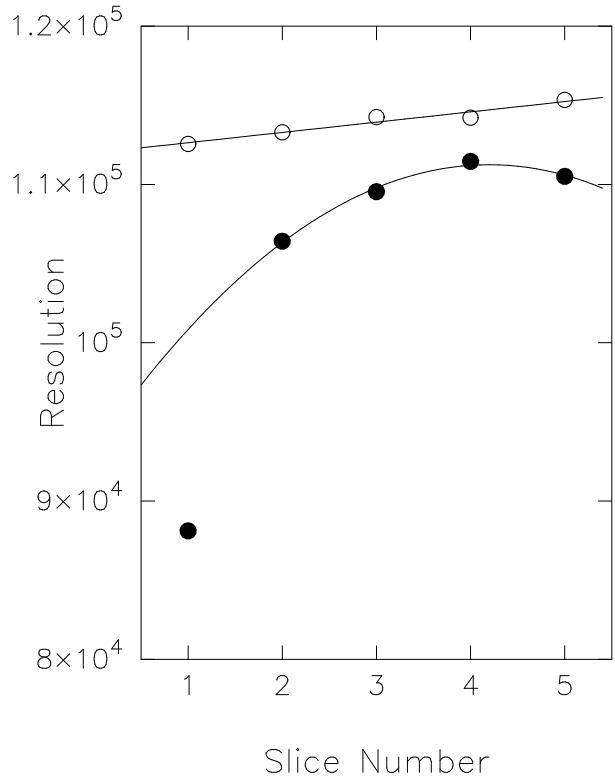

  \begin{center}
    \FigureFile(80mm,120mm){fig8.ps} 
  \end{center}
  \caption{The spectral resolution of each slice (number 1-5)
    obtained with the image slicer (filled circles) and with the slit of
    0.3 arcsec width (open circles). The parabola and linear fits are made
    for these data points. The first slice is excluded in the fit
    for the data obtained with the image slicer (see text).
} \label{fig:res}
\end{figure}

\subsection{Throughput}

\begin{figure}
  \begin{center}
    \FigureFile(80mm,80mm){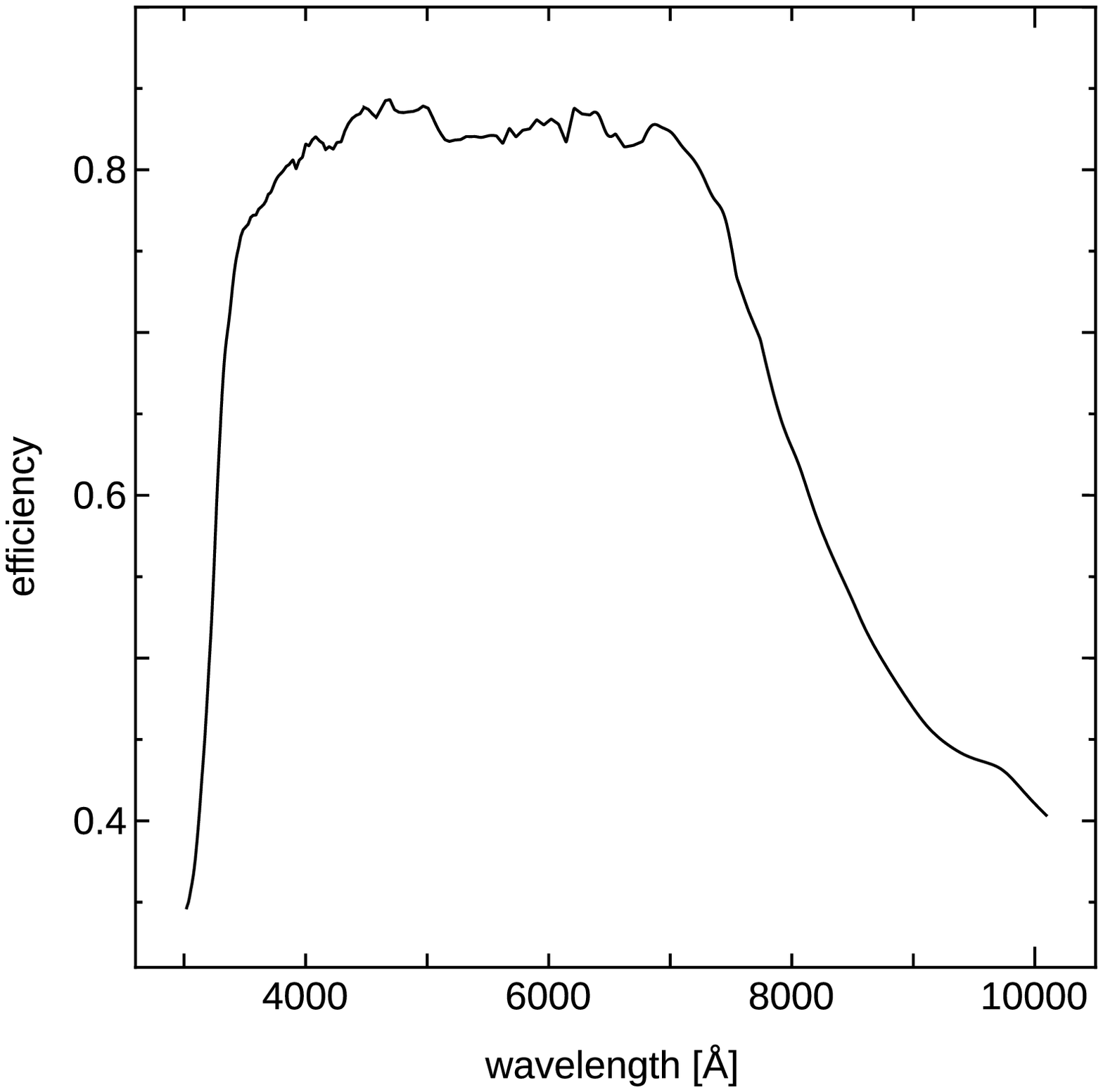} 
  \end{center}
  \caption{Throughput of the image slicer as a function of wavelength measured with the calibration lamp for flat-fielding.}\label{fig:throughput}
\end{figure}

The throughput of the image slicer is measured using the calibration
lamp for flat-fielding. The flat lamp is stable at the 0.1~\% level
\citep{tajitsu10}. The photon counts of the flat data obtained with
the image slicer are normalized by those obtained without the slicer.
Figure~\ref{fig:throughput} shows the result as a function of
wavelength. The measurements are made for three different setups with
different wavelength coverages. The count is measured at the center
of the spectral image for each echelle order.


The throughput of the image slicer is 80--85~\% in the wavelength
range from 400~nm to 700~nm, to which the AR coatings on the planes of
incidence and injection are optimized. The loss by the reflection at
the two surfaces of the slicer is expected to be as low as 1~\% in the
above wavelength range. The loss of light through the reflections
inside the slicer due to the absorption and/or scatter by small
particles would be as large as 10~\%. This can be estimated from the
decrease of the flat light from the first to fifth slices (see
figure~\ref{fig:cut}). Another potential source of loss is the scatter at
the slicer edges. Though the scatter is expected to be smaller than
5~\% from the manufacturing errors of the edges ($<10$~$\mu$m), the
estimate could be uncertain. The total throughput expected from the
above estimate is 85--90~\%.  Although the measured throughput is
slightly lower than this estimate, the instrument significantly
improves the efficiency of the observations (see
\S~\ref{sec:advantage}).


\subsection{Comparison of stellar spectra obtained with and without the image slicer}

\begin{figure}
  \begin{center}
    \FigureFile(80mm,80mm){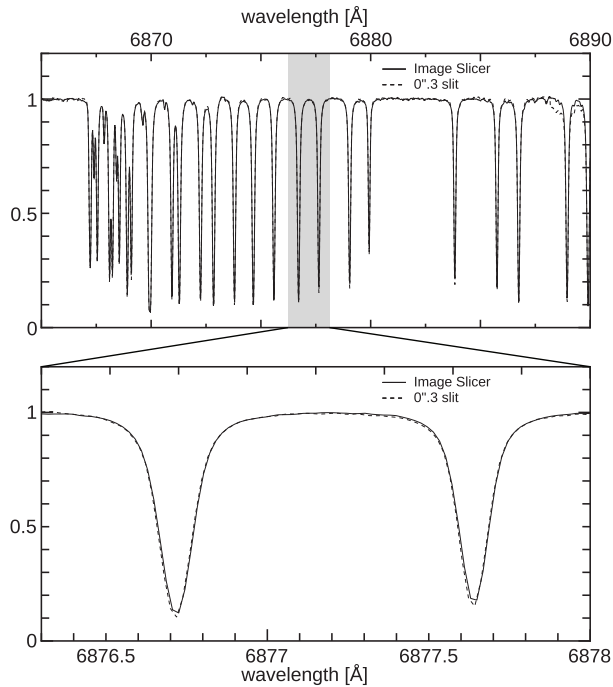} 
  \end{center}
  \caption{Comparison of stellar spectra obtained with and without the image slicer shown by the solid and dashed lines.}\label{fig:star}
\end{figure}

We obtained spectra of bright stars to compare the spectra obtained
using the image slicer with those obtained with the 0.3~arcsec
slit. Figure~\ref{fig:star} shows the spectra of o Per observed with
the S/N ratios of 530 (with the image slicer) and 400 (without the
image slicer) at 6860~{\AA}. The seeing when the both spectra were taken was
0.68 arcsec. The spectra obtained from the five individual slices
were combined.  Many sharp absorption lines of O$_{2}$
molecules in the Earth's atmosphere appear in this wavelength
range. We note that the continuum normalization is uncertain in the
range where many absorption lines are overlapping. However, the same
normalization procedure with the IRAF task ``continuum'' is applied to
both spectra with and without the image slicer, which basically cancel
the uncertainty in the comparison of the two spectra.

As expected from the comparisons of the Th-Ar emission lines
(\S~\ref{sec:th}), no significant difference is found between the two
spectra. We note that the spectrum obtained from the first slice,
whose line profile is not as sharp as in other spectra
(\S~\ref{sec:th}) is included in the stellar spectrum. However, since
the contribution of the first slice to the final data is less than
5~\% (figure~\ref{fig:cut}), the effect is negligible in the
comparison. However, in the event of poor seeing, in which the
fraction of light in the first slice is non-negligible, the light from
this slice may be ignored if high spectral resolution is paramount.

\section{Telescope setup, operations and data reduction}\label{sec:operation}

\subsection{Target acquisition and guiding}

\begin{figure}
  \begin{center}
    \FigureFile(120mm,120mm){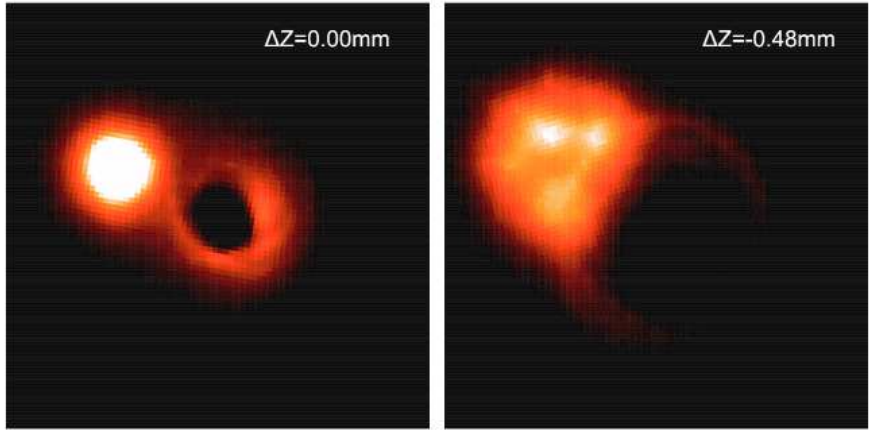} 
  \end{center}

  \caption{Slit viewer images for target acquisition and guiding. The
    left is the image obtained by focusing the telescope on the
    pinhole mirror in front of the image slicer. The target ($\rho$
    Oph) is on the pinhole of a 2.8~arcsec diameter at the mirror
    (center of the image). The companion star, 3 arcsec away from
    $\rho$ Oph, also appears in this image.  The right is the image
    after focusing the telescope on the image slicer. The guiding is
    made for such off-focus images. As seen in this image, objects
    around the target within several arcsec could result in
    contamination.  }
\label{fig:guide}
\end{figure}

The target acquisition is made using slit viewer images. The slit
viewer camera for the HDS is applied with no modification to the image
reflected by the plane mirror in front of the slicer. The target is
centered on the pinhole opened on the mirror that introduces the
stellar light to the image slicer (figure~\ref{fig:guide}). Since the
plane mirror is 6.97~mm in front of the slicer on which the
telescope should focus during observations, off-focus images are
acquired with the slit viewer camera for guiding. However, the
resulting image size of about 1.0--1.5~arcsec is sufficient to
guide the target to the pinhole, because targets of this
instrument observed with very high resolution are assumed to be bright
objects.

\subsection{Telescope focusing}

Since the position of the image slicer is different from that of the
original HDS slit, some offset of the telescope focus position is
necessary and this is achived by shifting the secondary mirror. There
is no facility to directly measure the image size at the position of
the slicer. Hence, the telescope focusing is performed using the slit
viewer images. The secondary mirror is then positioned with an offset
that was empirically determined to maximize the photon counts at the
central (third) slice.

\subsection{Advantages of the image slicer}\label{sec:advantage}

\begin{figure}
  \begin{center}
    \FigureFile(80mm,80mm){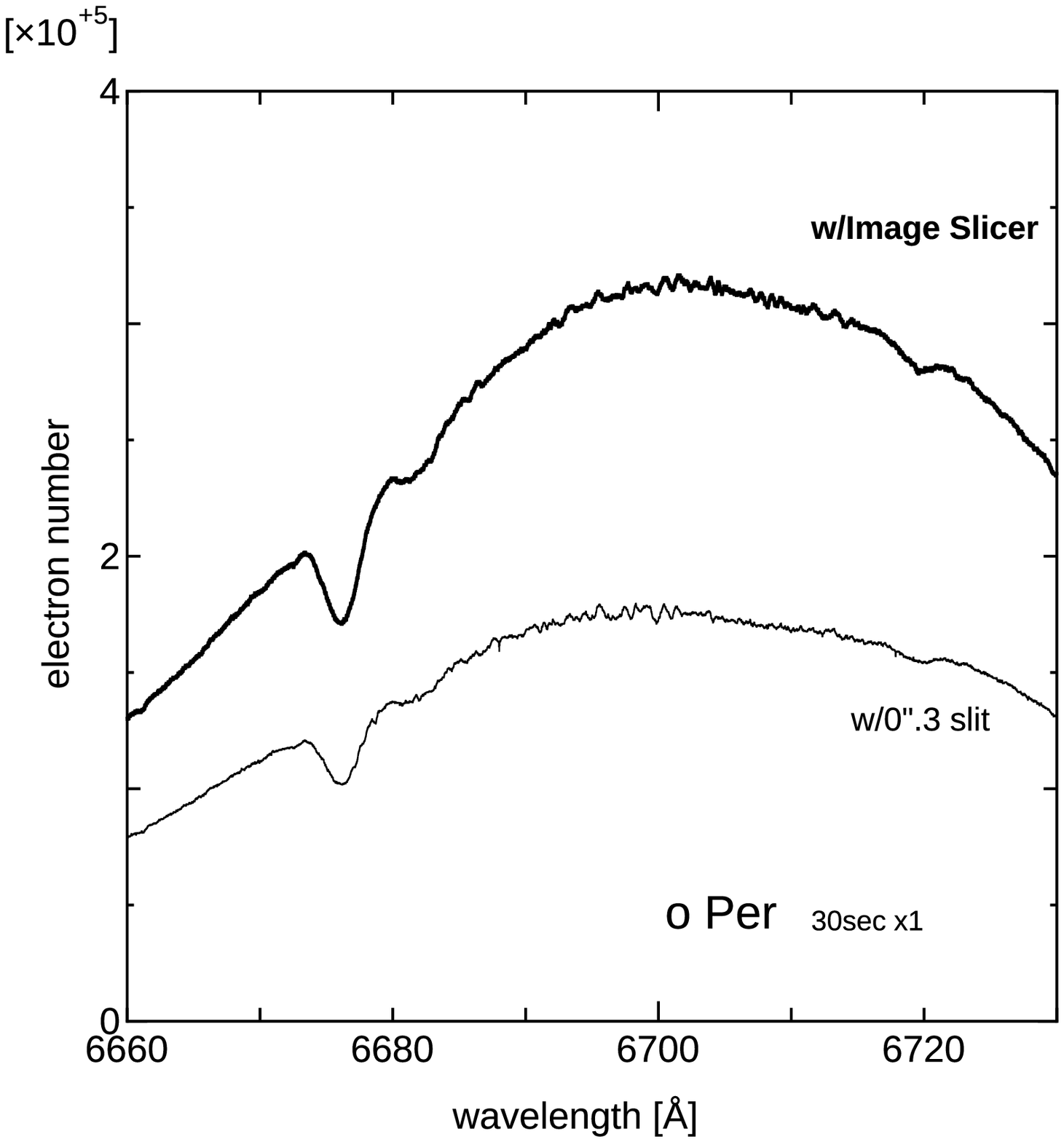} 
  \end{center}
  \caption{A comparison of spectra of o Per obtained with and without
    the image slicer. Each spectrum is flat-fielded and
    wavelength-calibrated, but is not normalized. The photon counts
    obtained with the slicer are 1.8 times higher than those with
    normal slit with the same exposure time, under the 0.68~arcsec
    seeing condition.}\label{fig:realobs}
\end{figure}

An example of comparisons of spectra obtained with and without the slicer
for the same exposure time is shown in figure~\ref{fig:realobs}. The
observation was taken in 0.68~arcsec seeing. The photon counts of
the spectrum obtained with the slicer are approximately 1.8 times
higher than that obtained with the normal slit. 

\begin{figure}
  \begin{center}
    \FigureFile(80mm,80mm){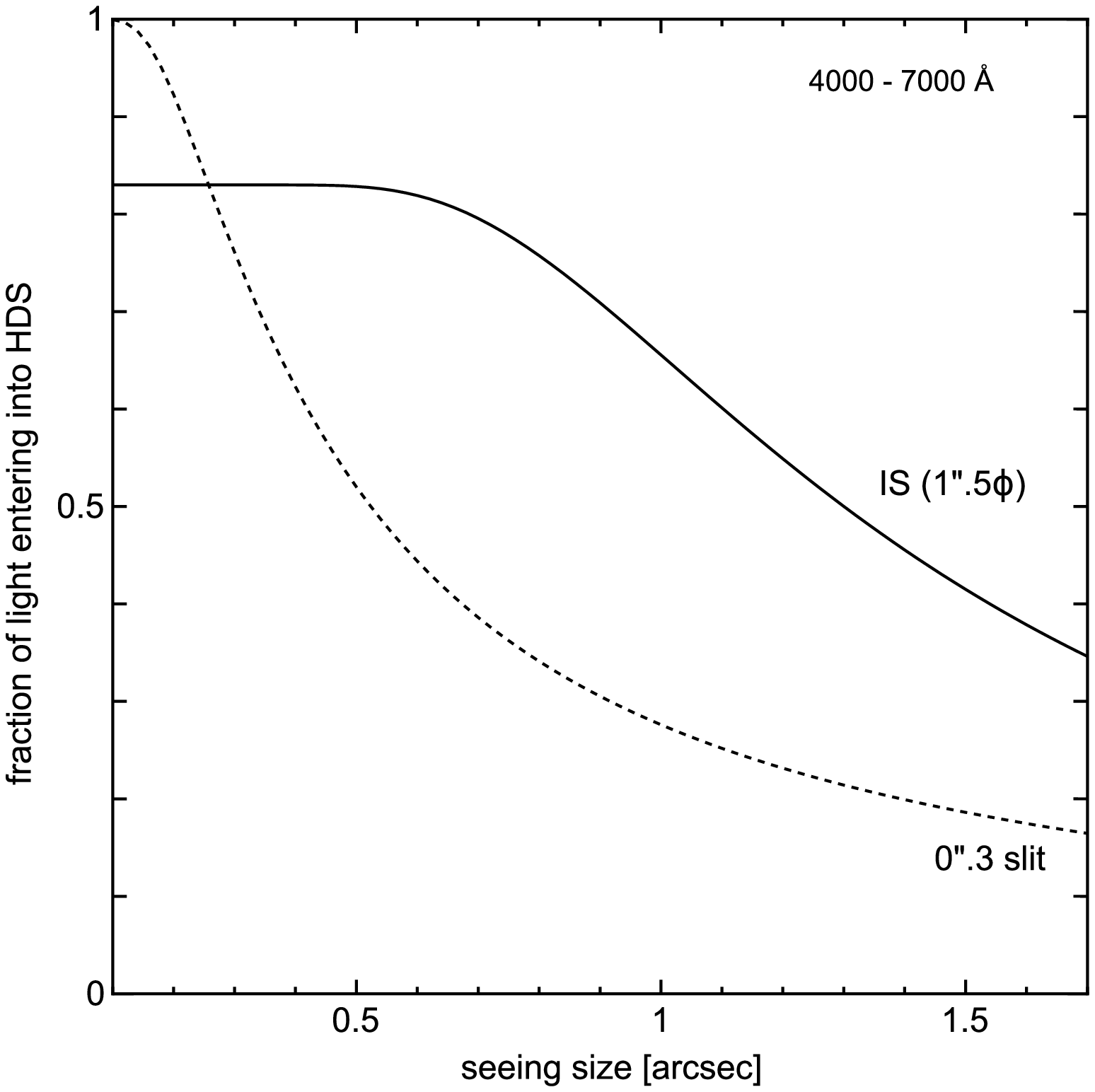} 
  \end{center}
  \caption{Fraction of light that enters into the spectrograph through
    the 0.3 arcsec slit (dotted line) and that through the image slicer (solid
    line)}\label{fig:efficiency}
\end{figure}

Figure~\ref{fig:efficiency} shows estimates of the fraction of
light that enters the spectrograph through the slit or the image
slicer as a function of the seeing size. The dotted line indicates the
value for the 0.3 arcsec slit, while the solid line shows the same for
the image slicer.


The gain of photons expected by using the image slicer, that is, the
ratio of the fraction of light entering to the spectrograph with the
image slicer to that for the 0.3 arcsec slit, is higher than unity for
the seeing size larger than 0.3 arcsec, and reaches to two at 0.8
arcsec. This is confirmed by observations for some cases, as shown in
figure~\ref{fig:realobs}. Hence, to obtain spectra with the resolution
as high as $R=110,000$, a higher S/N is expected by using the image
slicer in any usual seeing condition at Mauna Kea.

On the other hand, the spectrum with the image slicer spreads over
much number of CCD pixels along the slit length direction.  The
resultant spectrum is affected by the increase of the dark current,
the readout noise and the sky background. Moreover, the telescope
guiding and the target acquisition by the off-focus image (section
4.1) of faint targets could be difficult during actual
observations. Given these disadvantages for faint targets, the image
slicer is useful for objects brighter than 15 magnitudes.


\subsection{Data reduction technique}

In order to obtain the best quality data, the spectra should be
extracted separately for the five slices, for which wavelength
calibration is made using comparison Th-Ar spectra obtained by the
same procedure for the object data. Special care is required 
at the edge of the flat image (left- and right-hand sides of
the fifth and first slices of the flat data: see
figure~\ref{fig:cut}), because the regions of the detector illuminated
by the flat lamp are sometimes smaller than for those of an object.

The best quality spectrum is obtained by combining the four spectra of
2-5th slices after wavelength calibration.  However, the resolving
power of $R\sim 110,000$ is well achieved by combining the four
spectra before wavelength calibration, which significantly reduces the
reduction procedure.  The first slice might be used to increase the
photon counts, but that could result in a small decrease of spectral
resolution.

\section{Summary}

The design, manufacturing, and performance of the image slicer for the
Subaru/HDS are reported. Such an instrument enables one to obtain very
high resolution spectra with high efficiency. The instrument is
  already available for the common-use of HDS. A similar instrument
is also installed in the spectrograph of the 1.88~m telescope at
Okayama Astrophysical Observatory (OAO/HIDES: Izumiura et al. 1999),
which contributes to increasing the efficiency of the observations
with high resolving powers.

The image slicer reported here is designed to obtain a very high
resolving power ($R\sim 110,000$). We are also planning to install
another image slicer that provides smaller number of wider slice
images (e.g. three slices of 0.5 arcsec) to increase the efficiency of
observations with more usual spectral resolution ($R\sim
80,000$). The new image slicer will be installed in mid-2012
and also be opened for common-use.  Such efficient image slicers
will be more useful for next generations of larger telescopes.


\bigskip

The construction of the image slicer was supported by the Grant-in-Aid
for Science Research from JSPS (grant 20244035).

%
%


\end{document}